\documentclass{article}
\usepackage{amsmath}
\usepackage{amsfonts}

\usepackage[super,square,sort&compress,numbers]{natbib}
\newcommand{\eqdef}{\overset{\mathrm{def}}{=\joinrel=}}

\usepackage{graphicx}

\title{Vector quantisation and partitioning of COVID-19 temporal dynamics in the United States}
\author{Chris von Csefalvay\thanks{Starschema Inc., Arlington, VA. Correspondence: \texttt{csefalvayk@starschema.net}.}}

\begin{document}

\maketitle

\begin{abstract}
    The statistical dynamics of a pathogen within a population depend on a range of factors: population density, the effectiveness and investment into social distancing, public policy measures and non-pharmaceutical interventions (NPIs) are only some examples of factors that influence the number of cases over time by state. This paper outlines an analysis of time series vector quantisation and paritioning of COVID-19 cases in the United States, using a soft-DTW (Dynamic Time Warping) k-means clustering and a k-shape based clustering algorithm to identify internally consistent clusters of case counts over time. The identification of characteristic types of time-dependent variations can lead to the identification of patterns within sets of time series. This, in turn, can help discern the future of infectious dynamics in an area and, through identifying the most likely cluster-wise trajectory by calculating the cluster barycenter, inform public health decision-making.
\end{abstract}

\section{Introduction} 
\label{sec:introduction}

The emergence of SARS-CoV-2, and its associated viral syndrome COVID-19, has raised important questions about the ways we analyse and identify dynamic temporal processes. In particular, by identifying similarities in principal time-dependent indicators of epidemic dynamics, such as cumulative prevalence (the running total of confirmed cases over time), we can gain insight into similarities that are likely to emerge across various regions. Such similarities may be reflective of various hidden processes, be they related to the pathogen, to the response thereto or to various predisposing factors. By way of this, time series clustering has the potential to play a significant role in understanding the spatio-temporal factors governing the dynamic processes that drive an outbreak.

Vector quantisation and partitioning, or clustering, is the wider set of algorithms within unsupervised statistical learning that identify similar patterns among data in arbitrarily high-dimensional vector spaces, effectively taking a set $\mathcal{P}$ of $N$ vectors in an $n$-dimensional vector space and assigning to each of these a label from the label set $\mathcal{L}$, so that the assignment of each element of $\mathcal{P}$ to the groups defined by the labels comprising $\mathcal{L}$ minimise some objective function (typically referred to as the distance metric of the clustering). Cluster algorithms are widely used today and their practical applications are manifold, ranging from identifying clinical phenotypes in  medicine and population health\cite{ahmad2014clinical,haldar2008cluster,lochner2005cluster,weatherall2009distinct,ye2014different} through fraud detection\cite{behera2015credit,liu2013healthcare,peng2006application,sabau2012survey,subudhi2017use} to image segmentation.\cite{chuang2006fuzzy,coleman1979image,jin2018accelerating,lafata2018data,pappas1989adaptive,wu1993optimal}

Time series clustering presents a particular complication of this problem insofar as the subject of clustering is not a vector representing a single value, but rather a time series. These time series are typically not in synchrony, but rather exhibit a range of delays, lags and leads, and may depend on extrinsic and/or hidden variables. We may formulate the essential task of time series clustering as follows. Let $\mathcal{T}$ be a set of $n$ time series $x_{1 \ldots t_{max}}$. Further, let $k$ denote the cardinality of the label set $\mathcal{L}$ -- in other words, the number of partitions we wish to split the data into, with $k \leq n$. Then, the mapping $f: \mathcal{T} \rightarrow l \ | \ l \in \mathcal{L}$ is a clustering of the set of time series if it assigns to any element $x_i \in X \ | \ i \leq n$ one (and only one) label $l_i \in \mathcal{L}$, so as to minimise an objective function (typically referred to in this context as a distance metric) $J$ within each cluster $C_j = {x_p \in \mathcal{T} \ | \ l_p = j, j \in \mathcal{L}}$ defined by its label.

This paper examines the use of two time series clustering algorithms -- soft-DTW k-means clustering and k-shape clustering -- to identify different patterns in COVID-19 prevalence in the continental United States, and comparing the results of the classifiers for inter-classifier consistency. By isolating the barycenters of the time-shifted clusters, we can identify consistent patterns in prevalence dynamics across multiple states. This in turn can be used to quantify the overall effect of pre-existing characteristics, population dynamics and non-pharmaceutical interventions (NPIs) between states.


\section{Methods} 
\label{sec:methods}

\subsection{Source data} 
\label{sub:source_data}

\begin{figure}
	\includegraphics[width=\linewidth]{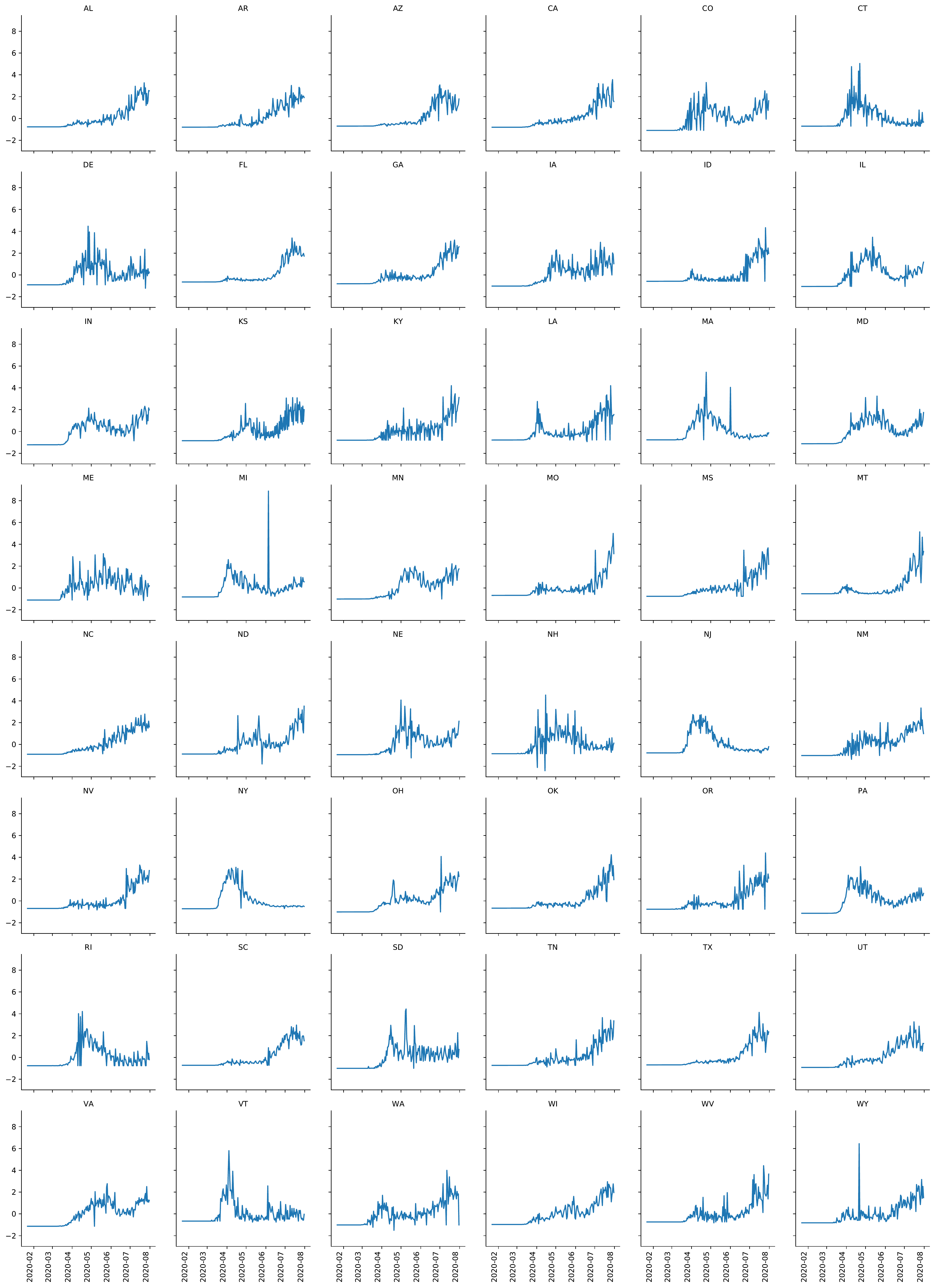}
	\centering
	\caption{Scaled raw data of prevalence by state.}
	\label{fig:scaled_by_state}
\end{figure}

Source data for the 48 states of the continental United States was obtained from the Starschema COVID-19 Data Set,\cite{foldi_tamas_2020_3969287} and filtered only for confirmed case counts. Data was loaded into Python 3.7 using \texttt{pandas},\cite{mckinney2011pandas} and values were scaled using \texttt{tslearn.preprocessing}'s \texttt{TimeSeriesScalerMeanVariance} to $\mu = 0$ and $\sigma = 1$. The results of this transformed raw data set are laid out, by state, in Figure~\ref{fig:scaled_by_state}. 

\subsection{Soft-DTW k-means clustering} 
\label{sub:soft_dtw_k_means_clustering}

Since first described by Sakoe and Chiba (1978),\cite{sakoe1978dynamic} the dynamic time warping algorithm has been expressed in multiple formulations. The presentation below is based on Cuturi and Blondel's 2017 paper introducing Soft-DTW, with the marginal difference of using $J(\cdot, \cdot)$ instead of $\delta$ to represent the distance function.\cite{cuturi2017soft} 

Given two time series $\mathbf{x}: t_x \in \mathbb{Z}$ and $\mathbf{y}: t_y \in \mathbb{Z}$, there exists a cost matrix $\Delta(\mathbf{x}, \mathbf{y})$ for the distance function $J$, from which we can derive the cost matrix 

\begin{equation}
	\Delta(\mathbf{x}, \mathbf{y}) = [J(x_i, y_j)]_{ij} \in \mathbb{R}^{t_x \times t_y} 
\end{equation}

For the two above-mentioned series, we may describe the set of matrices of all possible alignments as $\mathcal{A}_{t_x, t_y}$, which is a strict subset of $\{0, 1\}^{t_x \times t_y}$. Then, DTW can be defined as the function that for any pair of time series $(\mathbf{x}, \mathbf{y})$ identifies an alignment $A \in \mathcal{A}_{t_x, t_y}$ so as to minimise the inner product of A with the cost matrix $\Delta(\mathbf{x}, \mathbf{y})$ as

\begin{equation}
	DTW(\mathbf{x}, \mathbf{y}) \eqdef \min_{A \in \mathcal{A}_{t_x, t_y}} \langle A_{t_x, t_y}, \Delta(\mathbf{x}, \mathbf{y})
\end{equation}

Thus, DTW can be conceived of as a search task, in which $\mathcal{A}_{t_x, t_y}$ is the search space within which we search for an alignment $A$ given $\mathbf{x}$ and $\mathbf{y}$ so as to minimise the inner product $\langle A, \Delta(\mathbf{x}, \mathbf{y}) \rangle$. 

Soft-DTW universalises the notion underlying the DTW cost metric and the global alignment kernel metric

\begin{equation}
	GAK_{\gamma}(\mathbf{x}, \mathbf{y}) \eqdef \sum_{A \in \mathcal{A}_{t_x, t_y}} e^{- \frac{\langle A, \Delta(\mathbf{x}, \mathbf{y} \rangle)}{\gamma}}
	\label{eq:gak}
\end{equation}

\noindent into a single metric.\cite{janati2020spatio} Given the generalisation of the minimum metric with a smoothing factor $\gamma \geq 0$ as

\begin{equation}
	\min^{\gamma} \{a_{1 \ldots n}\} \eqdef 
	\begin{cases}
		\displaystyle \min_{i \leq n} a_i, 								& \gamma = 0 \\
		- \gamma \log \displaystyle \sum_{i=1}^n e^{-\frac{a_i}{\gamma}},	& \gamma > 0
	\end{cases}
\end{equation}

\noindent we may now define Soft-DTW as

\begin{equation}
	sDTW_{\gamma}(\mathbf{x}, \mathbf{y}) \eqdef \min_{A \in \mathcal{A}_{t_x, t_y}} ^{\gamma} \{ \langle A, \Delta(\mathbf{x}, \mathbf{y}) \rangle \}
\end{equation}

Importantly, Soft-DTW -- unlike the original DTW approach by Sakoe and Chiba\cite{sakoe1978dynamic} -- is explicitly differentiable. In particular, as Saigo (2006) noted,\cite{saigo2006optimizing} the gradient of 
 Equation~\eqref{eq:gak} can be calculated quite conveniently. Let $\hat{A}$ be the average alignment matrix following the Boltzmann distribution $p_{\gamma} \sim e^{- \langle A_i, \frac{\Delta(\mathbf{x}, \mathbf{y}) \rangle}{\gamma}}$ for all $A_i \in \mathcal{A}_{t_x, t_y}$. Then, 
 
\begin{equation}
 	\hat{A} = \frac{\displaystyle \sum_{A_i \in \mathcal{A}_{t_x, t_y}} A_i e^{- \langle A, \frac{\Delta(\mathbf{x}, \mathbf{y})}{\gamma}\rangle}}{GAK_{\gamma}(\mathbf{x}, \mathbf{y})}
 \end{equation}

\noindent and consequently

\begin{equation}
	\nabla_{\mathbf{x}} DTW_{\gamma} (\mathbf{x}, \mathbf{y}) = \Bigg( \frac{\partial \Delta(\mathbf{x}, \mathbf{y})}{\partial \mathbf{x}} \Bigg)^T \hat{A}
\end{equation}

This can be easily calculated using backward recursion, as described in Algorithm 2 of Cuturi and Blondel (2017).\cite{cuturi2017soft} In addition, the notion of a clustering centroid can be generalised to the metric space comprising the time series to yield Fr\^{e}chet means, also referred to in this context as barycenters. For a metric space $(M, \tau)$, $p \in M$ is a Fr\^{e}chet mean of order $q \geq 1$ of the time series $x_{1 \ldots n} \in M$ if it minimises the Fr\^{e}chet variance, i.e.

\begin{equation}
	p = \mathop{arg min}_{r \in M} \sum_{j = 1}^n \tau(x_j, r)^q \ | \ q \geq 1
\end{equation}

Thus, based on the soft-DTW metric, laid out above, we can extract from the time series of COVID-19 cumulative incidence a clustering that iteratively minimises the within-cluster sum of squares (k-means clustering). For the purposes of this paper, soft-DTW clustering was performed using \texttt{tslearn} 0.4.1\cite{JMLR:v21:20-091} using Python 3.7, with a $\gamma$ parameter of $0.1$.


\subsection{k-shape clustering} 
\label{sub:k_shape_clustering}

k-shape clustering is a novel, robust clustering algorithm for time series that relies on iteratively refining clusters, with cross-correlation as the underlying distance metric.\cite{paparrizos2015k} Specifically, k-shape relies on a normalised version of cross-correlation, referred to in this context as Shape Base Distance (SBD): time series are Z-normalised (i.e. $\mu = 0$ and $\sigma = 1$), and the resulting cross-correlation sequence is divided by the geometric mean of the individual time series' autocorrelations. In this sense, k-shape can be understood as a k-means clustering that uses a cross-correlation based metric $SBD(\mathbf{x}, \mathbf{y})$. Let $\mathbf{x}_s$ be the series $\mathbf{x}$ shifted, with zero-padding, by s, and the same be true for $\mathbf{y}_s$ respectively, \emph{mutatis mutandis}. For two time series of equal length $\mathbf{x}$ and $\mathbf{y}$, we recursively define shift-wise cross-correlation for shifts in the range $s \in [-m, m]$ as

\begin{equation}
	\psi_{k} (\mathbf{x}, \mathbf{y}) = 
	\begin{cases}
		\displaystyle \sum_{l = 1}^{m - k} x_{l + k} y_l 		& k \geq 0 \\
		\psi_{-k} (\mathbf{x}, \mathbf{y})						& k < 0 
	\end{cases}
\end{equation}

Then, for the cross-correlation sequence, we obtain the cross-correlation $\hat{\rho}_w$ for any value of $w \in {1, 2, \ldots, 2m - 1}$ as

\begin{equation}
	\hat{\rho}_w (\mathbf{x}, \mathbf{y}) = \psi_{w - m} (\mathbf{x}, \mathbf{y})
\end{equation}

\noindent Now, we can define the distance metric $SBD(\mathbf{x}, \mathbf{y})$ by

\begin{equation}
	SBD(\mathbf{x}, \mathbf{y}) = 1 - \max_w \Bigg( \frac{\hat{\rho_w} (\mathbf{x}, \mathbf{y})}{\sqrt{\psi_0(\mathbf{x}, \mathbf{x}) \cdot \psi_0(\mathbf{y}, \mathbf{y})}} \Bigg)
\end{equation}

Because of the convolution theorem, which states that under certain conditions convolution in one domain of a time series (or more generally, any signal) is equivalent to elementwise multiplication in the other domain,\cite{oppenheim2001discrete} we can efficiently compute $\psi(\mathbf{x}, \mathbf{y})$ by taking the complex conjugate of the discrete Fourier transform of each series $\mathcal{F}(\mathbf{x}) \star \mathcal{F}(\mathbf{y})$, where $\star$ is the complex conjugate operator.\cite{paparrizos2015k} Then, given the inverse discrete Fourier transform $\mathcal{F}^{-1}$,

\begin{equation}
	\psi(\mathbf{x}, \mathbf{y}) = \mathcal{F}^{-1} \Big( \mathcal{F}(\mathbf{x}) \star \mathcal{F}(\mathbf{y}) \Big)
\end{equation}

\noindent and as Paparrizzos and Gravano showed,\cite{paparrizos2015k} Fast Fourier Transforms allow this to be calculated efficiently in $\mathcal{O}(|\mathbf{x}| log(|\mathbf{x}|))$ time rather than $\mathcal{O}(|\mathbf{x}|^2)$ time.

Similarly to the cluster analysis carried out in Subsection~\ref{sub:soft_dtw_k_means_clustering}, k-shape clustering was performed using the \texttt{tslearn} package's \texttt{clustering.KShape} implementation, with an \texttt{n\_init} setting at 16 iterations for centroid seeds, using the result with the lowest inertia, random initialization and a convergence tolerance of $10^{-6}$.



\section{Results} 
\label{sec:results}

\subsection{Clustering time dynamics of disease prevalence} 
\label{sub:clustering_time_dynamics_of_disease_prevalence}

\begin{figure}
	\includegraphics[width=\linewidth]{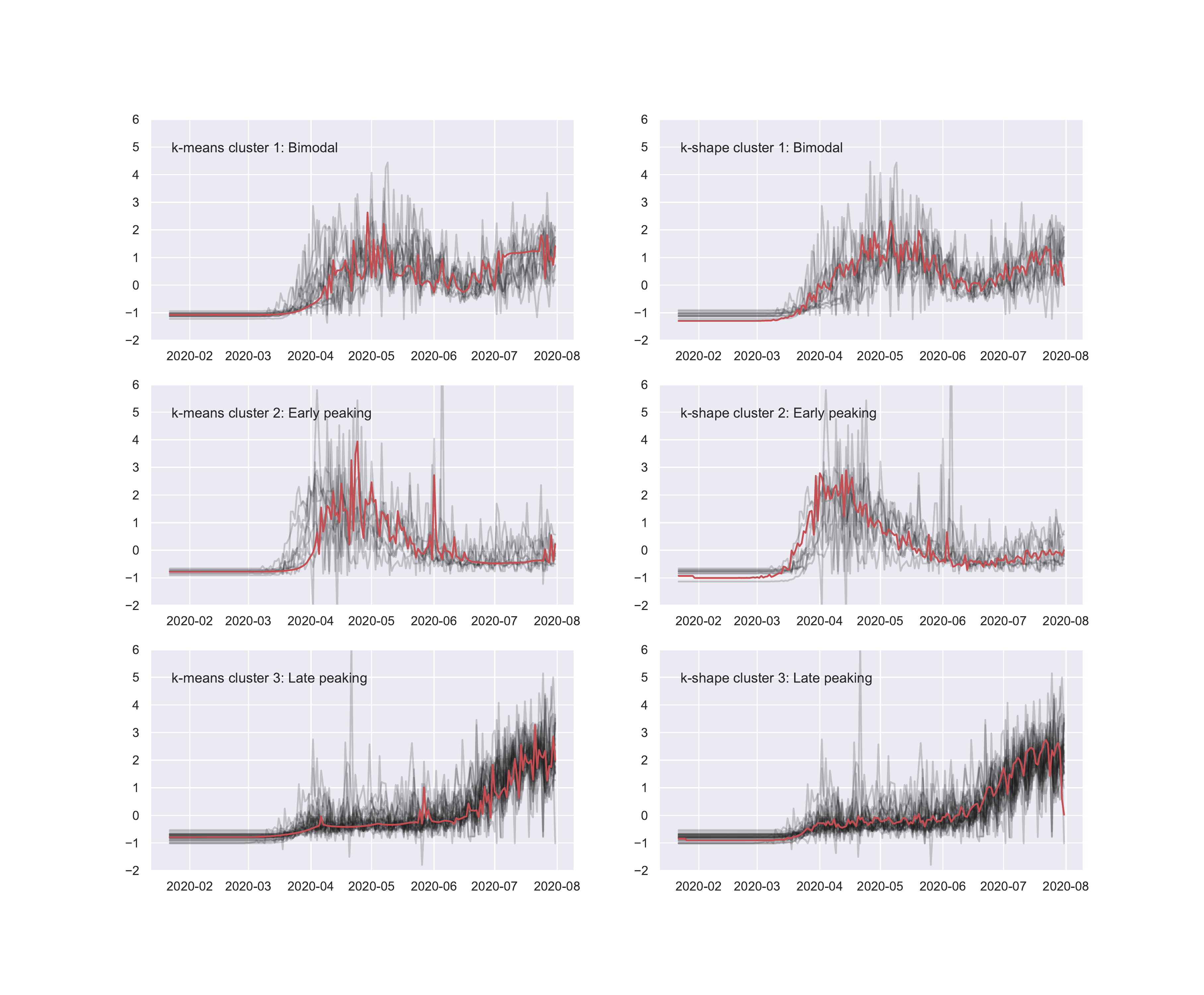}
	\centering
	\caption{Mutually consistent clusters (rows) between the k-means and k-shape cluster algorithms. Data is time adjusted and barycenters are displayed in red.}
	\label{fig:side-by-side}
\end{figure}

After fitting the soft-DTW k-means and k-shape clustering models on the data set described in Section~\ref{sub:source_data} with a label set cardinality (i.e. number of clusters) of 3, indicators for goodness of fit were obtained using \texttt{sklearn.metrics}. These show that the clustering is relatively sound. The silhouette scores (soft-DTW k-means: 0.249, k-shape: 0.276) indicate while there is some chance of an overlap, the clustering is a relatively good fit.\cite{rousseeuw1987silhouettes} This is confirmed by a strong Variance Ratio Criterion (Calinski-Harabasz score) of 18.809 and 20.155 for soft-DTW k-means and k-shape, respectively.\cite{calinski1974dendrite} As Figure~\ref{fig:side-by-side} clearly indicates, there are three distinctly characterisable patterns based on the barycenters:

\begin{enumerate}
	\item Late peaking (k-means cluster 1, k-shape cluster 1): states in this cluster typically have a steady, consistent pattern affected only by weekly periodicities, and begin to surge around mid-June 2020.
	\item Early peaking (k-means cluster 2, k-shape cluster 2): states in this cluster display a rapid-onset initial peak in April to May 2020, thereafter tapering off. 
	\item Bimodal (k-means cluster 3, k-shape cluster 3): within this cluster, states appear to exhibit a steady number of cases and the beginnings of a bimodal distribution over time, with a peak in April-May 2020 that subsides in June, then follows on to another rise in July and August.
\end{enumerate}

Figure~\ref{fig:side-by-side} highlights in red the barycenters or Fr\^{e}chet means of the time series, expanding the notion of a centroid as a central tendency to the metric space of the time series. While the barycenters are different between the clustering algorithms (largely due to small differences in clustering, thus leading to different compositions for the barycenter calculation), they identify consistently the underlying pattern characteristic of the cluster. Notably, the barycenters calculated by the k-shape classification exhibit much stronger short-term (weekly) periodicity in all three clusters. At the same time, the second cluster's abnormal peak in June is much less reflected in the barycenter based on the k-shape clustering than it is on the k-means cluster, and the k-shape cluster presents a barycenter with a much flatter 'peak' in mid-April than the k-means barycenter.

\begin{figure}
	\includegraphics[width=0.8\linewidth]{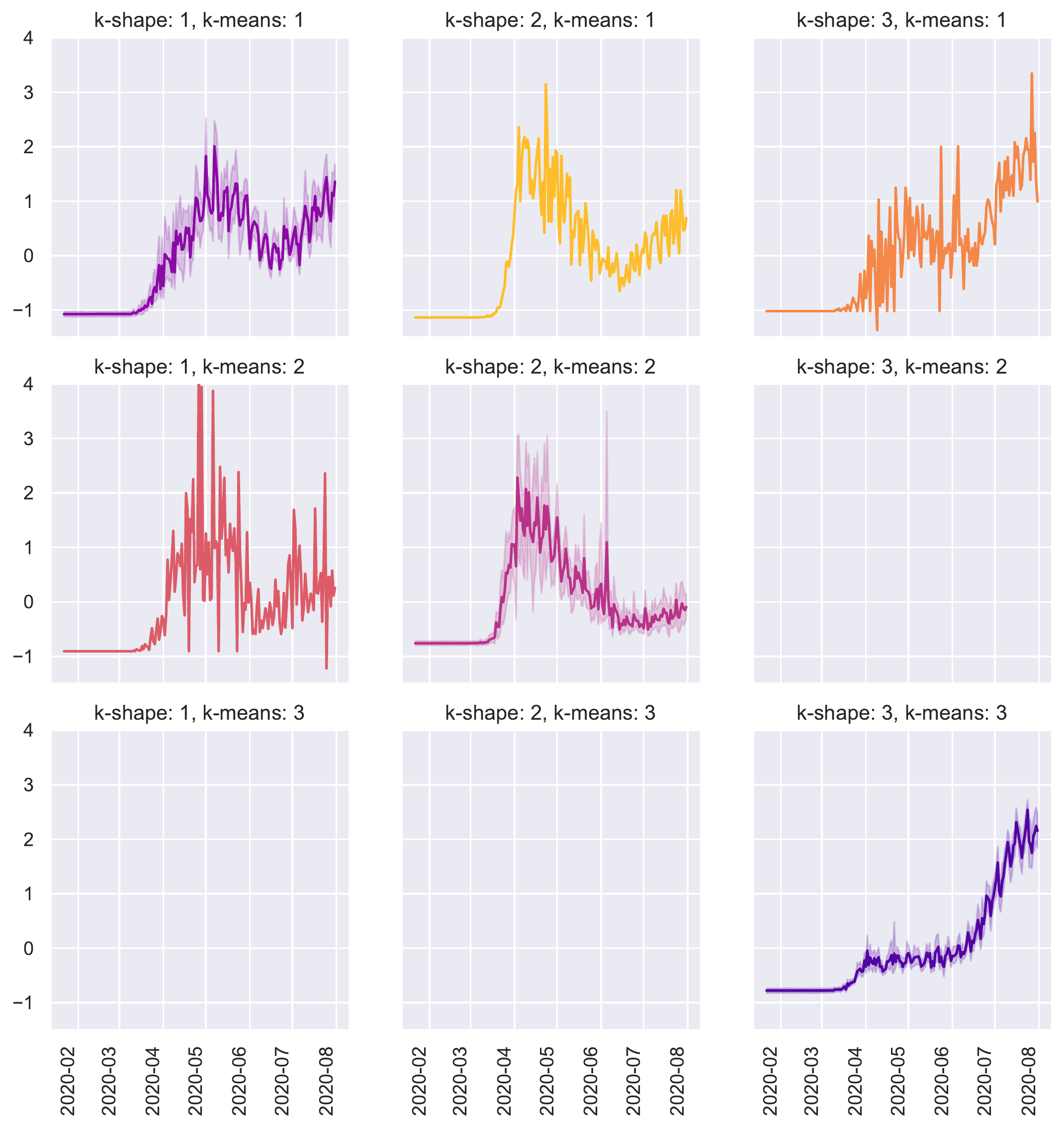}
	\centering
	\caption{Combined time traces of k-means and k-shape classifications for the major consensus groups. Bimodal behaviour accounts for 21\% of states, early-peaking behaviour covers 17\% and late-peaking, ascending behaviour accounts for over half (56\%) of states. Three states do not fall within the major consensus groups.}
	\label{fig:k_combinations}
\end{figure}

The distribution of time series (i.e. states) over the permutations of soft-DTW k-means and k-shape cluster assignments (see Figure~\ref{fig:k_combinations}) shows that the majority of states fall into matching soft-DTW k-means and k-shape categories, with only 3 states falling outside. Over half (56\%) of states fall into the soft-DTW k-means cluster 3 and k-shape cluster 3, while 8 states (17\%) fall into the soft-DTW k-means cluster 2 and k-shape cluster 2, and 9 states (21\%) fall into the soft-DTW k-means cluster 1 and k-shape cluster 1. This distribution is displayed in the inter-classifier agreement matrix in Figure~\ref{fig:inter_classifier}.


\subsection{Cross-cluster agreement} 
\label{sub:cross_cluster_agreement}

\begin{figure}
	\includegraphics[width=0.5\linewidth]{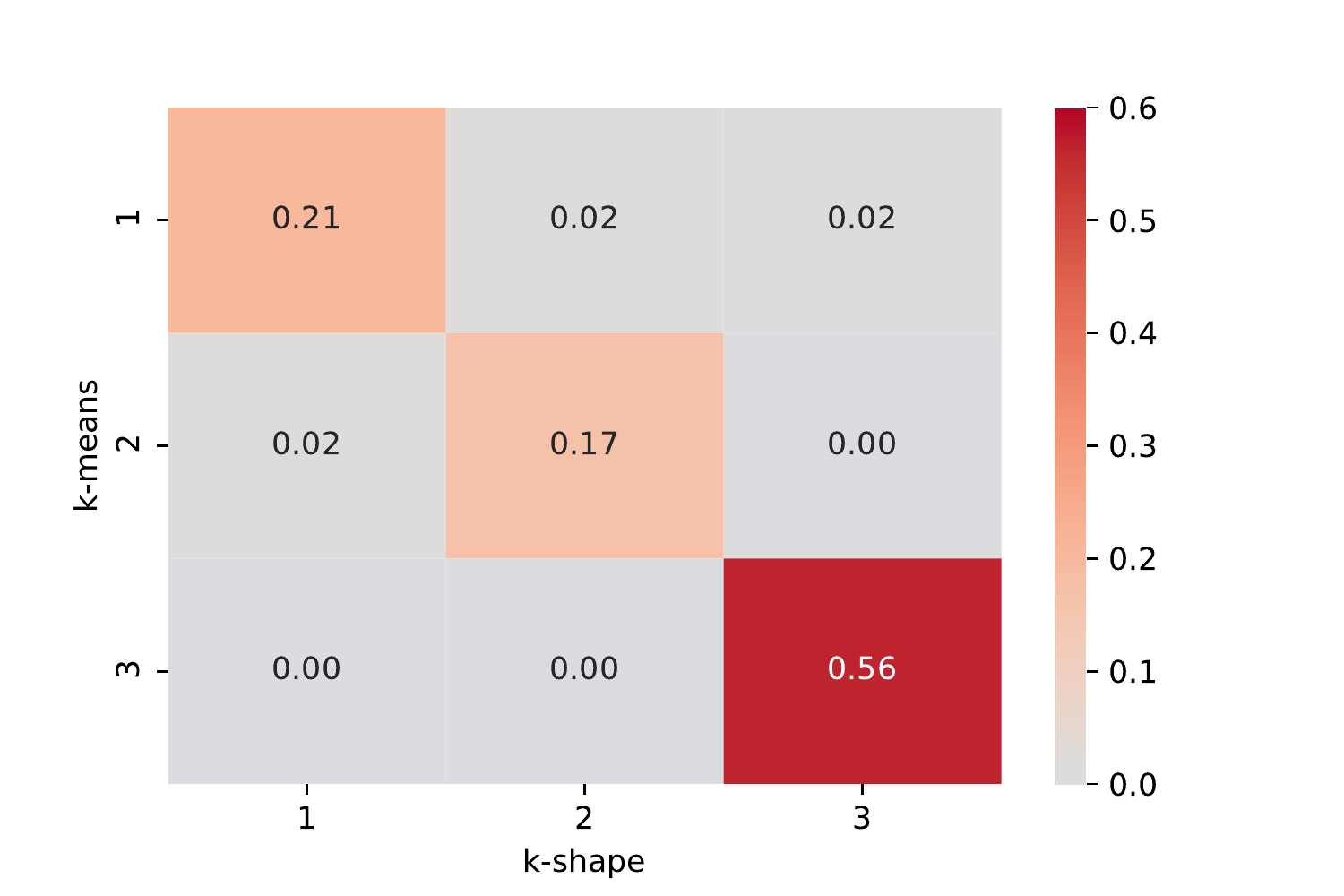}
	\centering
	\caption{Inter-classifier agreement between k-shape (\texttt{k-shape}) and soft-DTW k-means (\texttt{kmeans}) classification.}
	\label{fig:inter_classifier}
\end{figure}

In order to ascertain cross-cluster agreement, the Adjusted Rand Index (ARI) was used to quantify consensus between the k-shape and soft-DTW k-means classifiers.\cite{hubert1985comparing} This index, first proposed by Hubert and Arabie in 1985, is symmetric, thus it can be used to identify consensus between clusters with different metrics. At 0.864, the ARI indicates strong concurrence between the soft-DTW k-means and the k-shape classifiers.

Cross-cluster agreement is illustrated in Figure~\ref{fig:inter_classifier}. As it is evident therefrom, over half of the states fall into the late-peaking (k-means cluster 3, k-shape cluster 3) category, with relatively few cases and no pronounced peaks until June 2020, after which the data evidences an oscillating but gradually increasing case count. 

The strong cross-cluster agreement, covering 96\% of all samples, indicates that despite their methodological differences, both the soft-DTW k-means clustering algorithm and the k-shape algorithm yield largely identical results when it comes to assigning states' time series to clusters. The strong concurrence and favourable ARI indicate that the cluster assignments are unlikely to be artefactual results of the underlying algorithms but rather reflect truly significantly distinct groupings of states by their case count time series.



\section{Discussion} 
\label{sec:discussion}

k-shape and soft-DTW k-means classification strongly concur in identifying the three fundamental behavioural clusters of confirmed COVID-19 case count in the 48 states of the continental United States: a bimodal pattern, an early peaking pattern and a late, slower pattern that is largely stationary until approx. June 2020, then displays a rapid rise of cases.

\begin{figure}
	\includegraphics[width=\linewidth]{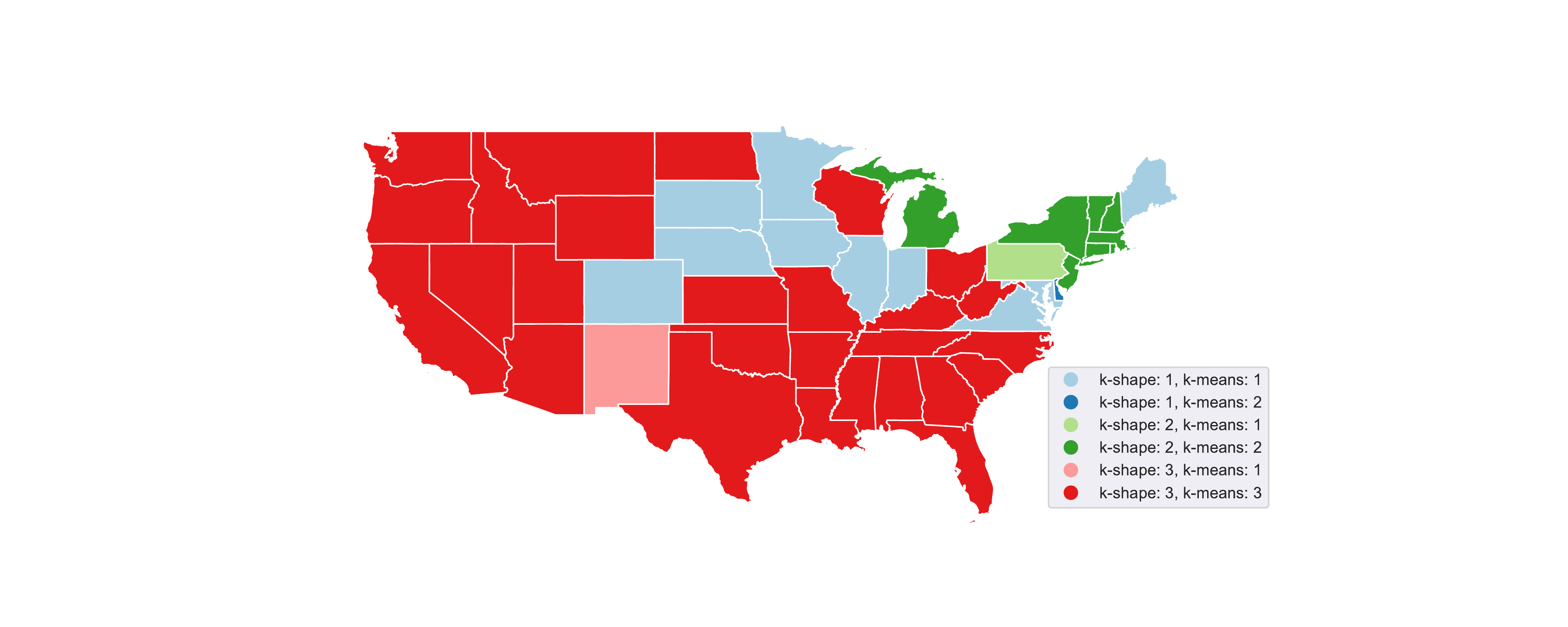}
	\centering
	\caption{Choropleth map of the United States displaying the permutations of k-shape and soft-DTW k-means clustering results by state.}
	\label{fig:choropleth}
\end{figure}

The geographical distribution of these is worth noting. As Figure~\ref{fig:choropleth} shows, at the time of writing, most of the area of the continental United States follows the late peaking regime, and the calculated barycenters indicate these states are currently poised to experience further growth in case counts. Only a few states (green shades) have followed an early outbreak with a significant reduction in cases and no further resurgence, as may be considered evidence of successful mitigation/suppression efforts on their part. Finally, a number of states (blue shades) have experienced early outbreaks and are exhibiting a bimodal pattern, whereby an initial surge in April to late May 2020 has been followed not by successful suppression but a reduction followed by yet another rise in the number of reported cases of COVID-19.

As this paper has shown, time series clustering allows for finding commonalities between time series that are by necessity out of synchrony. In doing so, it can be helpful in illuminating geographical and regional patterns of disease dynamics. In particular, by using two different methods -- a soft-DTW based, time-shifted k-means classifier and the correlation-based k-shape classifier --, a significant consensus between such classifications has been demonstrated where the number of confirmed COVID-19 cases in the continental United States is concerned. This lends credence to the hypothesis that epidemic dynamics of COVID-19 follow three distinct temporal patterns. These are in all likelihood conditioned by a combination of spatio-temporal factors (position along the epidemic's 'wavefront'), mitigation measures such as NPIs, their reltive effectiveness, as well as pre-existing factors of resilience and vulnerability.

Thus, by identifying the case count response, we can recognise different internally consistent clusters of case count progression over time. This may assist in understanding the governing patterns and dynamics of the SARS-CoV-2 pandemic, and assist in tailoring responses to the needs of individual areas and communities based on the temporal patterns of epidemic dynamics they exhibit.


\section*{Competing interests} 
\label{sec:competing_interests}

The author declares no competing interests.


\section*{Supplementary data} 
\label{sec:supplementary_data}

All simulations, code and data are available on Github and under the DOI \texttt{10.5281/zenodo.3970209}. Shape files for the choropleth diagram in Figure~\ref{fig:choropleth} have been obtained from the United States Census Bureau, and are included in the data set noted above.


\bibliography{bibliography}

\begin{thebibliography}{28}
\providecommand{\natexlab}[1]{#1}
\providecommand{\url}[1]{\texttt{#1}}
\expandafter\ifx\csname urlstyle\endcsname\relax
  \providecommand{\doi}[1]{doi: #1}\else
  \providecommand{\doi}{doi: \begingroup \urlstyle{rm}\Url}\fi

\bibitem[Ahmad et~al.(2014)Ahmad, Pencina, Schulte, O'Brien, Whellan, Pi{\~n}a,
  Kitzman, Lee, O'Connor, and Felker]{ahmad2014clinical}
Tariq Ahmad, Michael~J Pencina, Phillip~J Schulte, Emily O'Brien, David~J
  Whellan, Ileana~L Pi{\~n}a, Dalane~W Kitzman, Kerry~L Lee, Christopher~M
  O'Connor, and G~Michael Felker.
\newblock Clinical implications of chronic heart failure phenotypes defined by
  cluster analysis.
\newblock \emph{Journal of the American College of Cardiology}, 64\penalty0
  (17):\penalty0 1765--1774, 2014.

\bibitem[Haldar et~al.(2008)Haldar, Pavord, Shaw, Berry, Thomas, Brightling,
  Wardlaw, and Green]{haldar2008cluster}
Pranab Haldar, Ian~D Pavord, Dominic~E Shaw, Michael~A Berry, Michael Thomas,
  Christopher~E Brightling, Andrew~J Wardlaw, and Ruth~H Green.
\newblock Cluster analysis and clinical asthma phenotypes.
\newblock \emph{American {J}ournal of {R}espiratory and {C}ritical {C}are
  {M}edicine}, 178\penalty0 (3):\penalty0 218--224, 2008.

\bibitem[Lochner et~al.(2005)Lochner, Hemmings, Kinnear, Niehaus, Nel,
  Corfield, Moolman-Smook, Seedat, and Stein]{lochner2005cluster}
Christine Lochner, Sian~MJ Hemmings, Craig~J Kinnear, Dana~JH Niehaus, Daniel~G
  Nel, Valerie~A Corfield, Johanna~C Moolman-Smook, Soraya Seedat, and Dan~J
  Stein.
\newblock Cluster analysis of obsessive-compulsive spectrum disorders in
  patients with obsessive-compulsive disorder: clinical and genetic correlates.
\newblock \emph{Comprehensive {P}sychiatry}, 46\penalty0 (1):\penalty0 14--19,
  2005.

\bibitem[Weatherall et~al.(2009)Weatherall, Travers, Shirtcliffe, Marsh,
  Williams, Nowitz, Aldington, and Beasley]{weatherall2009distinct}
M~Weatherall, J~Travers, PM~Shirtcliffe, SE~Marsh, MV~Williams, MR~Nowitz,
  S~Aldington, and R~Beasley.
\newblock Distinct clinical phenotypes of airways disease defined by cluster
  analysis.
\newblock \emph{European Respiratory Journal}, 34\penalty0 (4):\penalty0
  812--818, 2009.

\bibitem[Ye et~al.(2014)Ye, Pien, Ratcliffe, Bj{\"o}rnsdottir, Arnardottir,
  Pack, Benediktsdottir, and Gislason]{ye2014different}
Lichuan Ye, Grace~W Pien, Sarah~J Ratcliffe, Erla Bj{\"o}rnsdottir, Erna~Sif
  Arnardottir, Allan~I Pack, Bryndis Benediktsdottir, and Thorarinn Gislason.
\newblock The different clinical faces of obstructive sleep apnoea: a cluster
  analysis.
\newblock \emph{European Respiratory Journal}, 44\penalty0 (6):\penalty0
  1600--1607, 2014.

\bibitem[Behera and Panigrahi(2015)]{behera2015credit}
Tanmay~Kumar Behera and Suvasini Panigrahi.
\newblock Credit card fraud detection: a hybrid approach using fuzzy clustering
  \& neural network.
\newblock In \emph{2015 Second International Conference on Advances in
  Computing and Communication Engineering}, pages 494--499. IEEE, 2015.

\bibitem[Liu and Vasarhelyi(2013)]{liu2013healthcare}
Qi~Liu and Miklos Vasarhelyi.
\newblock Healthcare fraud detection: A survey and a clustering model
  incorporating geo-location information.
\newblock In \emph{29th {W}orld {C}ontinuous {A}uditing and {R}eporting
  {S}ymposium (29WCARS), Brisbane, Australia}, 2013.

\bibitem[Peng et~al.(2006)Peng, Kou, Sabatka, Chen, Khazanchi, and
  Shi]{peng2006application}
Yi~Peng, Gang Kou, Alan Sabatka, Zhengxin Chen, Deepak Khazanchi, and Yong Shi.
\newblock Application of clustering methods to health insurance fraud
  detection.
\newblock In \emph{2006 International Conference on Service Systems and Service
  Management}, volume~1, pages 116--120. IEEE, 2006.

\bibitem[Sabau(2012)]{sabau2012survey}
Andrei~Sorin Sabau.
\newblock Survey of clustering based financial fraud detection research.
\newblock \emph{Informatica Economica}, 16\penalty0 (1):\penalty0 110, 2012.

\bibitem[Subudhi and Panigrahi(2017)]{subudhi2017use}
Sharmila Subudhi and Suvasini Panigrahi.
\newblock Use of optimized fuzzy c-means clustering and supervised classifiers
  for automobile insurance fraud detection.
\newblock \emph{Journal of King Saud University-Computer and Information
  Sciences}, 2017.

\bibitem[Chuang et~al.(2006)Chuang, Tzeng, Chen, Wu, and Chen]{chuang2006fuzzy}
Keh-Shih Chuang, Hong-Long Tzeng, Sharon Chen, Jay Wu, and Tzong-Jer Chen.
\newblock Fuzzy c-means clustering with spatial information for image
  segmentation.
\newblock \emph{{C}omputerized {M}edical {I}maging and {G}raphics}, 30\penalty0
  (1):\penalty0 9--15, 2006.

\bibitem[Coleman and Andrews(1979)]{coleman1979image}
Guy~Barrett Coleman and Harry~C Andrews.
\newblock Image segmentation by clustering.
\newblock \emph{Proceedings of the IEEE}, 67\penalty0 (5):\penalty0 773--785,
  1979.

\bibitem[Jin et~al.(2018)Jin, Xie, Huang, and Hussain]{jin2018accelerating}
Xiao-Bo Jin, Guo-Sen Xie, Kaizhu Huang, and Amir Hussain.
\newblock Accelerating infinite ensemble of clustering by pivot features.
\newblock \emph{Cognitive Computation}, 10\penalty0 (6):\penalty0 1042--1050,
  2018.

\bibitem[Lafata et~al.(2018)Lafata, Zhou, Liu, and Yin]{lafata2018data}
Kyle Lafata, Zhennan Zhou, Jian-Guo Liu, and Fang-Fang Yin.
\newblock Data clustering based on {Langevin} annealing with a self-consistent
  potential.
\newblock \emph{arXiv preprint arXiv:1806.10597}, 2018.

\bibitem[Pappas and Jayant(1989)]{pappas1989adaptive}
Thrasyvoulos~N Pappas and Nikil~S Jayant.
\newblock An adaptive clustering algorithm for image segmentation.
\newblock In \emph{International Conference on Acoustics, Speech, and Signal
  Processing}, pages 1667--1670. IEEE, 1989.

\bibitem[Wu and Leahy(1993)]{wu1993optimal}
Zhenyu Wu and Richard Leahy.
\newblock An optimal graph theoretic approach to data clustering: Theory and
  its application to image segmentation.
\newblock \emph{IEEE {T}ransactions on {P}attern {A}nalysis and {M}achine
  {I}ntelligence}, 15\penalty0 (11):\penalty0 1101--1113, 1993.

\bibitem[Tam{\'a}s and von Csefalvay(2020)]{foldi_tamas_2020_3969287}
F{\"o}ldi Tam{\'a}s and Chris von Csefalvay.
\newblock Starschema covid-19 data set, August 2020.
\newblock URL \url{https://doi.org/10.5281/zenodo.3969287}.

\bibitem[McKinney et~al.(2011)]{mckinney2011pandas}
Wes McKinney et~al.
\newblock pandas: a foundational python library for data analysis and
  statistics.
\newblock \emph{Python for High Performance and Scientific Computing},
  14\penalty0 (9), 2011.

\bibitem[Sakoe and Chiba(1978)]{sakoe1978dynamic}
Hiroaki Sakoe and Seibi Chiba.
\newblock Dynamic programming algorithm optimization for spoken word
  recognition.
\newblock \emph{IEEE {T}ransactions on {A}coustics, {S}peech, and {S}ignal
  {P}rocessing}, 26\penalty0 (1):\penalty0 43--49, 1978.

\bibitem[Cuturi and Blondel(2017)]{cuturi2017soft}
Marco Cuturi and Mathieu Blondel.
\newblock Soft-dtw: a differentiable loss function for time-series.
\newblock \emph{arXiv preprint arXiv:1703.01541}, 2017.

\bibitem[Janati et~al.(2020)Janati, Cuturi, and Gramfort]{janati2020spatio}
Hicham Janati, Marco Cuturi, and Alexandre Gramfort.
\newblock Spatio-temporal alignments: Optimal transport through space and time.
\newblock In \emph{International Conference on Artificial Intelligence and
  Statistics}, pages 1695--1704, 2020.

\bibitem[Saigo et~al.(2006)Saigo, Vert, and Akutsu]{saigo2006optimizing}
Hiroto Saigo, Jean-Philippe Vert, and Tatsuya Akutsu.
\newblock Optimizing amino acid substitution matrices with a local alignment
  kernel.
\newblock \emph{BMC {B}ioinformatics}, 7\penalty0 (1):\penalty0 246, 2006.

\bibitem[Tavenard et~al.(2020)Tavenard, Faouzi, Vandewiele, Divo, Androz,
  Holtz, Payne, Yurchak, Ru{\ss}wurm, Kolar, and Woods]{JMLR:v21:20-091}
Romain Tavenard, Johann Faouzi, Gilles Vandewiele, Felix Divo, Guillaume
  Androz, Chester Holtz, Marie Payne, Roman Yurchak, Marc Ru{\ss}wurm, Kushal
  Kolar, and Eli Woods.
\newblock Tslearn, a machine learning toolkit for time series data.
\newblock \emph{Journal of Machine Learning Research}, 21\penalty0
  (118):\penalty0 1--6, 2020.
\newblock URL \url{http://jmlr.org/papers/v21/20-091.html}.

\bibitem[Paparrizos and Gravano(2015)]{paparrizos2015k}
John Paparrizos and Luis Gravano.
\newblock k-shape: Efficient and accurate clustering of time series.
\newblock In \emph{Proceedings of the 2015 ACM SIGMOD International Conference
  on Management of Data}, pages 1855--1870, 2015.

\bibitem[Oppenheim et~al.(2001)Oppenheim, Buck, and
  Schafer]{oppenheim2001discrete}
Alan~V Oppenheim, John~R Buck, and Ronald~W Schafer.
\newblock \emph{Discrete-time signal processing. Vol. 2}.
\newblock Upper Saddle River, NJ: Prentice Hall, 2001.

\bibitem[Rousseeuw(1987)]{rousseeuw1987silhouettes}
Peter~J Rousseeuw.
\newblock Silhouettes: a graphical aid to the interpretation and validation of
  cluster analysis.
\newblock \emph{Journal of {C}omputational and {A}pplied {M}athematics},
  20:\penalty0 53--65, 1987.

\bibitem[Cali{\'n}ski and Harabasz(1974)]{calinski1974dendrite}
Tadeusz Cali{\'n}ski and Jerzy Harabasz.
\newblock A dendrite method for cluster analysis.
\newblock \emph{Communications in {S}tatistics -- {T}heory and {M}ethods},
  3\penalty0 (1):\penalty0 1--27, 1974.

\bibitem[Hubert and Arabie(1985)]{hubert1985comparing}
Lawrence Hubert and Phipps Arabie.
\newblock Comparing partitions.
\newblock \emph{Journal of {C}lassification}, 2\penalty0 (1):\penalty0
  193--218, 1985.

\end{thebibliography}

\end{document}